\RequirePackage{fix-cm}
\documentclass[pdftex,twocolumn,epjc3]{svjour3}  
\smartqed  
\RequirePackage{graphicx}

\usepackage{graphicx}%
\usepackage{subfigure}
\usepackage{multirow}%
\usepackage{amsmath,amssymb,amsfonts}%
\usepackage{amsthm}%
\usepackage{mathrsfs}%
\usepackage[title]{appendix}%
\usepackage{xcolor}%
\usepackage{textcomp}%
\usepackage{manyfoot}%
\usepackage{booktabs}%
\usepackage{algorithm}%
\usepackage{algorithmicx}%
\usepackage{algpseudocode}%
\usepackage{lineno}
\usepackage{url}
\RequirePackage{graphicx}
\RequirePackage{mathptmx}      
\RequirePackage{flushend}
\RequirePackage[numbers,sort&compress]{natbib}
\RequirePackage[colorlinks,citecolor=blue,urlcolor=blue,linkcolor=blue]{hyperref}

\graphicspath{{./}{figures/}}
\usepackage{longtable}

\newcommand{\bea}{\begin{eqnarray}}
\newcommand{\eea}{\end{eqnarray}}
 
\newcommand{\ep}{\epsilon}

\usepackage{threeparttable}

\begin{document}
%
%
%
%
\journalname{Eur. Phys. J. C}

\title{Constraining dark photon parameters based on the very high energy observations of blazars}


\author{Tian-Ci Liu$^{1,\rm {a}}$ 
        \and
        Ming-Xuan Lu$^{1,\rm {b}}$
        \and
        Xiao-Song Hu$^{1}$
}

\thankstext{e1}{e-mail: ltc@st.gxu.edu.cn}
\thankstext{e2}{e-mail: lumx@st.gxu.edu.cn}

\institute{Guangxi Key Laboratory for Relativistic Astrophysics,
	School of Physical Science and Technology, Guangxi University, Nanning 530004,
	China
}

\date{Received: date / Accepted: date}

\maketitle

\begin{abstract}
Dark photon is a new gauge boson beyond the Standard Model as a kind of dark matter (DM) candidate. Dark photon dark matter (DPDM) interacts with electromagnetic fields via kinetic mixing, implicating an approach to give a constraint with extragalactic very high energy (VHE) sources. 
In this work, we attempt to constrain the kinetic mixing from the photon-dark photon scattering process in the host galaxy of blazar, the intergalactic medium and the Milky Way.
The VHE photons from a blazar would pass through a dense DM spike around the supermassive black hole where the absorption from DPDM is dramatically enhanced. 
The kinetic mixing is constrained to be $\ep \sim 10^{-7}$ at a 95$\%$ confidence level with $m_{\rm D}\sim 0.03 - 1$ eV mass range from the observations of Markarian (Mrk) 421 and Mrk 501.
\end{abstract}

\section{Introduction}
\label{sect:intro}

Dark matter \cite[DM;][]{1996PhR...267..195J,2005PhR...405..279B}
dominates the matter density and constitutes 27$\%$ of the total energy density in our Universe. The existence of DM has been powerfully proved by a lot of evidence \cite{1987ARA&A..25..425T,2018RvMP...90d5002B}, but its nature still remains elusive because of the no result of direct detection \cite{2014PDU.....4...50B, 2017PhRvL.118b1303A} and indirect detection researches \cite{2022ChPhC..46c0005B,2017JHEP...01..107C, 2021PhRvD.103d3018B,2020JCAP...11..029A,2008JCAP...07..008R,2010PhLB..689..149E,2021PhLB..82136611C,2023arXiv230816762C,2007PhRvL..99w1102H,2022PhRvR...4a2022X,2023arXiv230906151H,2023arXiv231011391G}.

 The ordinary photon is indicated to be able to interact with low-mass bosonic states by some well-motivated scenarios \cite{1988PhRvD..37.1237R,1997NuPhB.492..104D,2008PhLB..666...66A}. We focus on the simplest case with only extra $U(1)$ symmetry corresponding to a new gauge boson, also known as dark photon \cite{2020arXiv200501515F,1986PhLB..166..196H,1980PhLB...95..285F,1990NuPhB.347..743F}. 
Dark photon could provide a possible explanation for some shortcomings of the Standard Model (SM) and also be a potential DM candidate as a very light vector boson.
The existence of dark photons has been predicted by some string-inspired models \cite{2008JHEP...07..073B,2009JHEP...11..027G,2011JHEP...07..114C}. 

Dark photons could mix the ordinary photons kinetically, inducing the oscillations and scattering processes between photons and dark photons \cite{2021PhRvD.103d3018B}.
This scattering process often leads to an observed flux reduction, which depends on the energy and propagating distance of the ordinary photons. The DM density where the photons propagate is also a vital aspect. 
Because of the benefits of the extragalactic very high energy (VHE) observations, we tentatively utilize the attenuation of gamma-ray survival probability induced by dark photons to place a boundary in its parameter space.

The extragalactic VHE sources are crucial astrophysical observational targets. Their photons will propagate several Mpc of distance scale before arriving at Earth.
However, the extragalactic VHE sources are rarely observed, especially when associated with transient events. 
Blazar Markarian (Mrk) 421 and Mrk 501 are famous extragalactic VHE sources where the photons of TeV energy scale have been detected \cite{2001A&A...366...62A,2011ApJ...738...25A}. Blazar is a type of active galactic nuclei (AGN) with the jet basically towards the Earth \cite{1995PASP..107..803U}. They are characterized by a nonthermal photon spectral energy distribution with two peaks, and the higher one at gamma-ray frequencies is our target. 

The DM profile near the supermassive black hole (SMBH) at the center of the blazar host galaxy is expected to form a spike with a sharper halo index if the SMBH grew adiabatically from an initial seed in the DM halo \cite{1999PhRvL..83.1719G,2022SCPMA..6569512L,2022PhRvL.128v1104W,2023PhRvL.130i1402C,2014PhRvL.113o1302F,2001PhRvD..64d3504U}. In that case, the DM density is obviously larger than the one expected from a naive extrapolation of the galactic density profile. Therefore, the absorption of VHE photons from the spike part is significant even if the spike region is only within $\sim 1$ pc.

The outline of this paper is as follows: In Section \ref{sect:scatter}, we briefly introduce the photon-dark photon scattering process and the absorption of ordinary photons caused.
In Section \ref{sect:profile}, the DM density where VHE photons propagate is discussed in detail, especially for the DM spike surrounding the SMBH of AGNs. The total DM mass along the line of sight (LOS) is also calculated to measure the effect of the spike with different emitting region sizes.
We talked about the constraining method in Section \ref{sect:method} and present the final results in Section \ref{sect:results}. Finally, we summarize and discuss the possibility of constraining the kinetic mixing by GRB221009A in Section \ref{sect:conclusion}.

\section{The photon-dark photon scattering}
\label{sect:scatter}

In extensions of the SM, the interaction of dark photons and the SM particles are described by the Lagrangian \cite{2019JCAP...11..020F}
\begin{equation}
	\mathcal{L}_{{\rm SM}\otimes {\rm D}}=-\epsilon e J_\mu^{\rm SM}A_{\rm D}^\mu\
\end{equation}
where $\ep$ is the kinetic mixing parameter that allows the nonzero mass dark photon to directly couple to the ordinary matter, $J_\mu^{\rm SM}$ is the SM electromagnetic current and $A_{\rm D}^\mu$ is the potential of dark photons with vector $U(\rm 1)$. $\ep$ could be as large as $\sim$ 1 theoretically since the vector portal is a dimension-four operator and unsuppressed by any high mass scale \cite{2013arXiv1311.0029E}. However, the minimum value of $\ep$ is still not clear and predicted to be $10^{-12}$ - $10^{-3}$ \cite{2008JHEP...07..124A,2009JHEP...11..027G,2012JHEP...01..021G,2011JHEP...07..114C}.

When the energies are above that of the dark photon mass but below the energies of the corresponding fermion mass, the above terms can be integrated out to yield the familiar low-energy interactions \cite{2021PhRvD.103d3018B}
\begin{equation}
{\cal L} \supset -\frac{1}{4}{\cal F}^{\mu\nu} {\cal F}_{\mu\nu} - \frac{\epsilon}{2} {\cal F}^{\mu\nu} {\cal F'}_{\mu\nu} -\frac{1}{4}{\cal F'}^{\mu\nu} {\cal F'}_{\mu\nu}\,.
\end{equation}
The scattering cross-section is represented as
\bea
\sigma_{\rm D} &=& \frac{8\pi\epsilon^2\alpha^2}{3(s-m_{\rm D}^2)^3}\left[-\beta(s^2+4sm_e^2+m_{\rm D}^4)\vphantom{\left(\frac{1+\beta}{1-\beta}\right)}\right.+\ln\left(\frac{1+\beta}{1-\beta}\right)\notag\\
 &&\left.(s^2+4sm_e^2+m_{\rm D}^4-4m_{\rm D}^2m_e^2-8m_e^4)\right]
\label{eq:cs}
\eea
where $\alpha$ is the fine structure constant, $m_e$ is the electron mass, $m_{\rm D}$ is the DPDM mass, $\beta = \sqrt{1-4m_e^2/s}\,$ and $s=2E m_{\rm D}+ m^2_{\rm D}$ are the usual Mandelstam variables.

The scattering process $\gamma \gamma'\rightarrow e^+ e^-$ becomes kinetically open up for ordinary photons above the energy threshold, which is given by
\begin{equation}
    E > E_{\rm th} = \frac{2m_e^2}{m_{\rm D}}.
\end{equation}
Dark photons of $m_{\rm D}$ = $10^{-2}$ eV could scatter with ordinary photons that reach 52 TeV energy threshold where the highest photon energy of Mrk 501 is only $\sim18.6$ TeV. Therefore, we use the $m_{\rm D}$ $\sim$ $10^{-2}$ eV as the minimum dark photon mass in the following calculation.

Assuming that the DM is composed of dark photons, the optical depth caused by the scattering with DPDM in the DM halo and IGM space could be calculated as
\begin{equation}
\label{tau}
\begin{aligned}
\begin{split}  
    \tau_{\rm DP} &=\tau{\rm _{host}} + \tau{\rm _{IGM}} + \tau{\rm _{MW}} \\  &= \int_{\rm los} (n_{\rm D, host}+n_{\rm D, IGM}+n_{\rm D, MW}) \sigma_{\rm D} {\rm d}l  
\end{split}
\end{aligned}
\end{equation}
where $l$ is the propagating distance of photons in the DM halo and IGM space along the LOS. $n_{\rm D} = \rho_{\rm DM}/m_{\rm D}$ is the DPDM number density ($n_{\rm D, host}$, $n_{\rm D, IGM}$ and $n_{\rm D, MW}$ denote that of the host galaxy, IGM space and the Milky Way respectively.) 
The ordinary photon survival probability induced by the photon-dark photon scattering is expressed as
\begin{equation}
    P_{\rm DP} = e^{-\tau_{\rm DP}}.
\end{equation}

The propagating directions are determined by the location of the VHE sources, indicating different propagating routes and DPDM number densities in the Milky Way halo toward the Earth. The information of Mrk 421 and Mrk 501 are shown in Table \ref{source information}.

\begin{table}
	\begin{center}
	\caption{The information of Mrk 421 and Mrk 501.}
	\label{source information}
	\renewcommand\arraystretch{1.5}
	\begin{tabular}{l|ccccc}
	\hline
	Source &  RA & DEC & $z$ & log($M_{\rm {BH}}/M_{\odot}$)\\
	\midrule
    Mrk 501  & 16:53:52.21 & 39:45:36.6 & 0.034 & 9.21 $\pm$ 0.13\\
    Mrk 421 & 11:04:27.31 & 38:12:31.7 & 0.031 & 8.28 $\pm$ 0.11\\
	\hline
	\end{tabular}
	\end{center}
    \begin{tablenotes}
	\item
     The coordinates are from: \href{http://tevcat2.uchicago.edu/tool}{http://tevcat2.uchicago.edu/tool}	
	\end{tablenotes}
\end{table}

\section{Dark matter profile}
\label{sect:profile}

The adiabatic growth of a SMBH at the central region of a galaxy is expected to form a sharper index of DM profile named spike, which was first discussed by Gondolo and Silk \cite{1999PhRvL..83.1719G}. 
The DM density inside the radius of the gravitational influence of the SMBH will be dramatically enhanced because of the dense DM spike. 
A spike around SMBH would arise if the majority of the SMBH growth from a seed occuring within the inner 50 pc of the DM cusp with a $10^7$ year timescale \cite{2001PhRvD..64d3504U}.

\begin{figure*}
\centering
\subfigure[$\gamma_{sp}$ = $\frac{9-2\gamma}{4-\gamma}$]{ 
\includegraphics[scale=0.45]{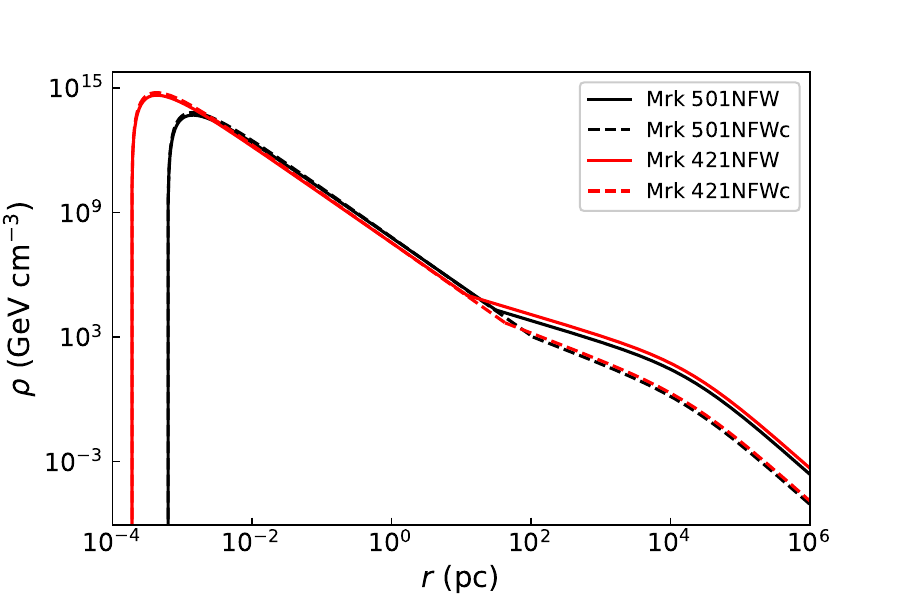}}
\subfigure[$\gamma_{sp}$ = 1.5]{
\includegraphics[scale=0.45]{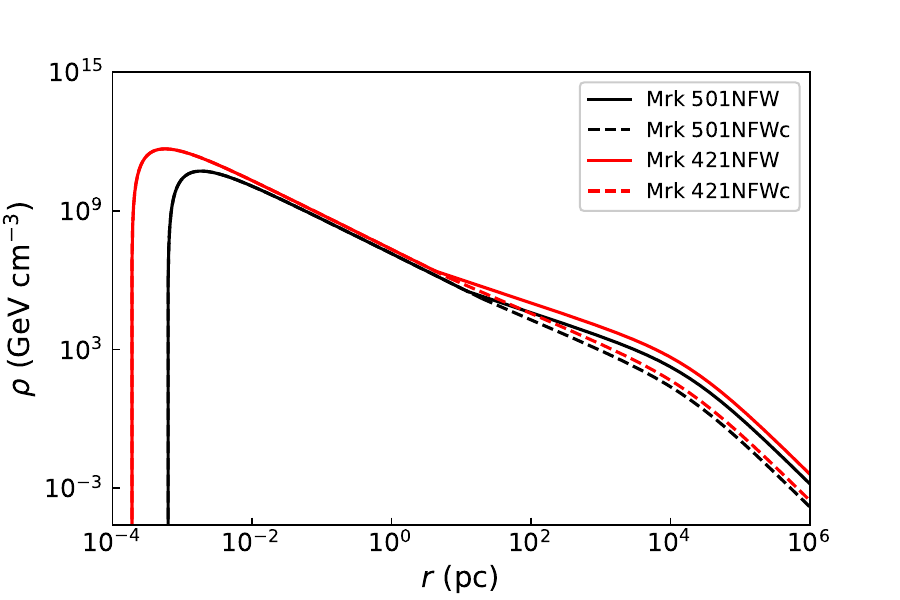}}
\caption{The DM profiles of Mrk 501 and Mrk 421 host galaxies for different $\gamma_{sp}$ based on NFW (solid lines) and NFWc (dashed lines) profiles.}
\label{DM_host_profile}
\end{figure*}

For annihilating DM particles, e.g. weakly interacting massive particles (WIMPs), a saturation density would be reached in the inner part of the spike. Since the DM particle is assumed as dark photon, the spike would form a simple power-law as \cite{2022SCPMA..6569512L,2023JCAP...05..057F}
\begin{equation}
    \rho_{\rm {sp}}(r) = \rho_{\rm {R}} g_{\gamma}(r)(\frac{R_{\rm {sp}}}{r})^{\gamma_{\rm {sp}}}
\end{equation}
where $\gamma_{\rm {sp}}$ = $\frac{9-2\gamma}{4-\gamma}$ is the cusp index of the DM profile within spike, $R_{\rm {sp}}$ = $\alpha r_0(M_{\rm {BH}}/(\rho_0 r_0^3))^{\frac{1}{3-\gamma}}$ is the size of spike with a numerical coefficient $\alpha \simeq$ 0.1 for ${\gamma}$ = 1 \cite{1999PhRvL..83.1719G}, $g_{\gamma}(r) \simeq (1 - \frac{4R_s}{r})^3$ and $R_s$ is the Schwarzschild radius of SMBH. 
 $\rho_R = \rho_0(R_{sp}/r_0)^{-\gamma}$ denotes a normalization factor chosen to align with the DM profile out of the spike region, where the outside profile is adopted as Navarro-Frenk-White (NFW) \cite{1996ApJ...462..563N} expressed as \cite{1998ApJ...502...48K} 
\begin{equation}
    \rho(r) = \frac{\rho_0}{(r/r_s)^\gamma(1+r/r_0)^{3-\gamma} }
\end{equation}
where $r_0$ and $\rho_0$ are the characteristic radius and density of the NFW profile respectively.
We assume that the DM distribution follows the canonical NFW and NFWc profile where $\gamma$ = 1 and 1.3. $r_0$ is fixed to be 20 kpc as for the Milky Way.

The normalization of DM spike at AGNs could be fixed by requiring the DM mass within the spike not exceed the uncertainty of the SMBH mass \cite{2010PhRvD..82h3514G,2017PhRvD..96f3008L},
\begin{equation}
    \int_{4R_s}^{10^5R_s}4\pi r^2\rho(r){\rm d}r \lesssim \Delta M_{\rm BH}
\label{normalization}
\end{equation}
where the SMBH masses and uncertainties are also shown in Table \ref{source information} \cite{2002ApJ...579..530W,2003ApJ...583..134B}, $R_{\rm min}$ is adopted as $4 R_s$ where the DM spike vanish because of the capture onto the SMBH, and $10^5 R_s$ is a typical size of the adiabatically contracted spike. The DM profiles of Mrk 421 and Mrk 501 based on NFW and NFWc are shown in Fig. \ref{DM_host_profile} left panel. It is obvious that the normalization of DM spike would hardly affected by the parameter $\gamma$. Thus we only consider the canonical NFW situation.

A conservative situation studied by Gnedin and Primack \cite{2004PhRvL..93f1302G} indicates that the DM particles scattering with the dense stellar cluster around the SMBH gravitationally would give a universal inner profile that scales as $r^{-3/2}$. In this case, the DM profiles are also shown in Fig. \ref{DM_host_profile} right panel.

Furthermore, the DM profile of the Milky Way is also adopted as NFW and $\rho_0$ is fixed by the local DM density $\rho\rm{_{DM}}$$ (r)$ = 0.4 GeV cm$^{-3}$ at $r$ = 8.5 kpc \cite{2010A&A...523A..83S,2011JCAP...11..029I,2021RPPh...84j4901D}.
Before the VHE photons arrive at the Earth, they have propagated a long path in the IGM space. This part could usually be regarded as a clean space. Based on the $\Lambda$CDM model, the DM density $\rho_{\rm{DM}}$ = $\Omega_{\rm {DM}}$ $\times$ $\rho_{c}$, where $\Omega_{\rm DM}$ $\sim$ 0.27 and $\rho_{c}$ = 3$H_{0}^{2}/8\pi G$
$\sim$ 8.6 $\times 10^{-30}$ g/cm$^3$ are the closure parameter and critical density of the universe today \cite{2020A&A...641A...6P}.

\begin{figure}
\centering
\includegraphics[scale=0.55]{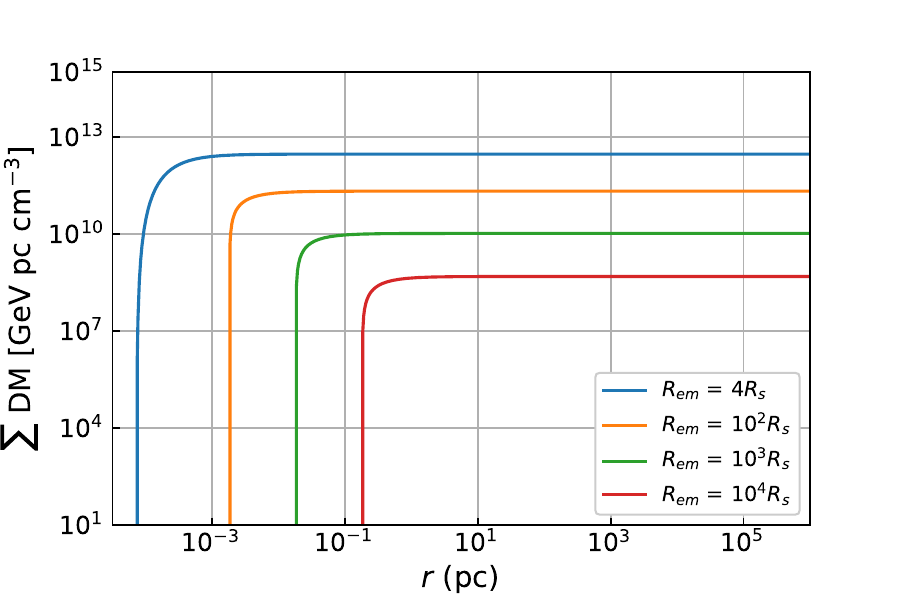}
\caption{The DM mass integral of Mrk 421 along the LOS to different emission region $R_{\rm em}$. }
\label{Mrk421_sum_DM}
\end{figure}

As mentioned in Chapter \ref{sect:scatter}, the optical depth $\tau_{\rm DP}$ could be calculated for each VHE source considering the DPDM in the IGM space, DM halo of host galaxy and the Milky Way. $\tau_{\rm DP}$ is dependent on the total DM mass along the LOS to the emission region of the VHE photon with a fixed scattering cross-section. For Mrk421 and Mrk 501, the extremes values of
emitting region size up to $10^3-10^4$ $R_s$ are not ruled out
\cite{2011ApJ...736..131A,2011ApJ...727..129A}.
The DM mass integral with emission region $R_{\rm em}$ = $4 R_s$, $100 R_s$ ,$10^3 R_s$ and $10^4 R_s$ of e.g. Mrk 421, are shown as Fig.
\ref{Mrk421_sum_DM} \cite{2022PhRvL.128v1104W}.

\begin{table*}
	\begin{center}
	\caption{The best-fit parameters of the null hypothesis as the ECPL model.}
	\label{without DP result}
	\renewcommand\arraystretch{1.5}
	\begin{tabular}{l|cccc}
	\hline
	Source & A & $\alpha$ & $E_{\rm {cut}}$ & $\chi^{2}/dof$\\
        &($10^{-10}$TeV$^{-1}$cm$^{-2}$s$^{-1}$)&&TeV&\\
	\midrule
    Mrk 421 & $1.51^{+0.03}_{-0.03}$ & 2.27$^{+0.07}_{-0.07}$ & $4.85^{+0.64}_{-0.51}$ & 24.39/10\\
    Mrk 501 & $1.38^{+0.03}_{-0.03}$ & 1.93$^{+0.03}_{-0.04}$ & $14.60^{+2.48}_{-2.02}$ & 47.23/23\\
	\hline
	\end{tabular}
	\end{center}
\end{table*}

\begin{figure*}
\centering
\subfigure[Mrk 421]{
\includegraphics[scale=0.45]{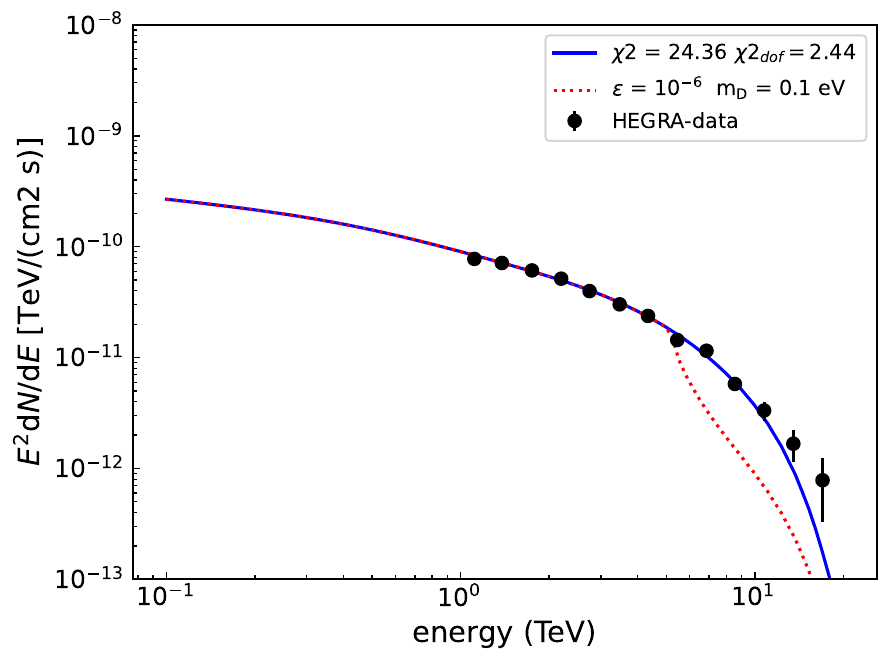}}
\subfigure[The posterior distribution of ECPL parameters for Mrk 421]{
\includegraphics[scale=0.27]{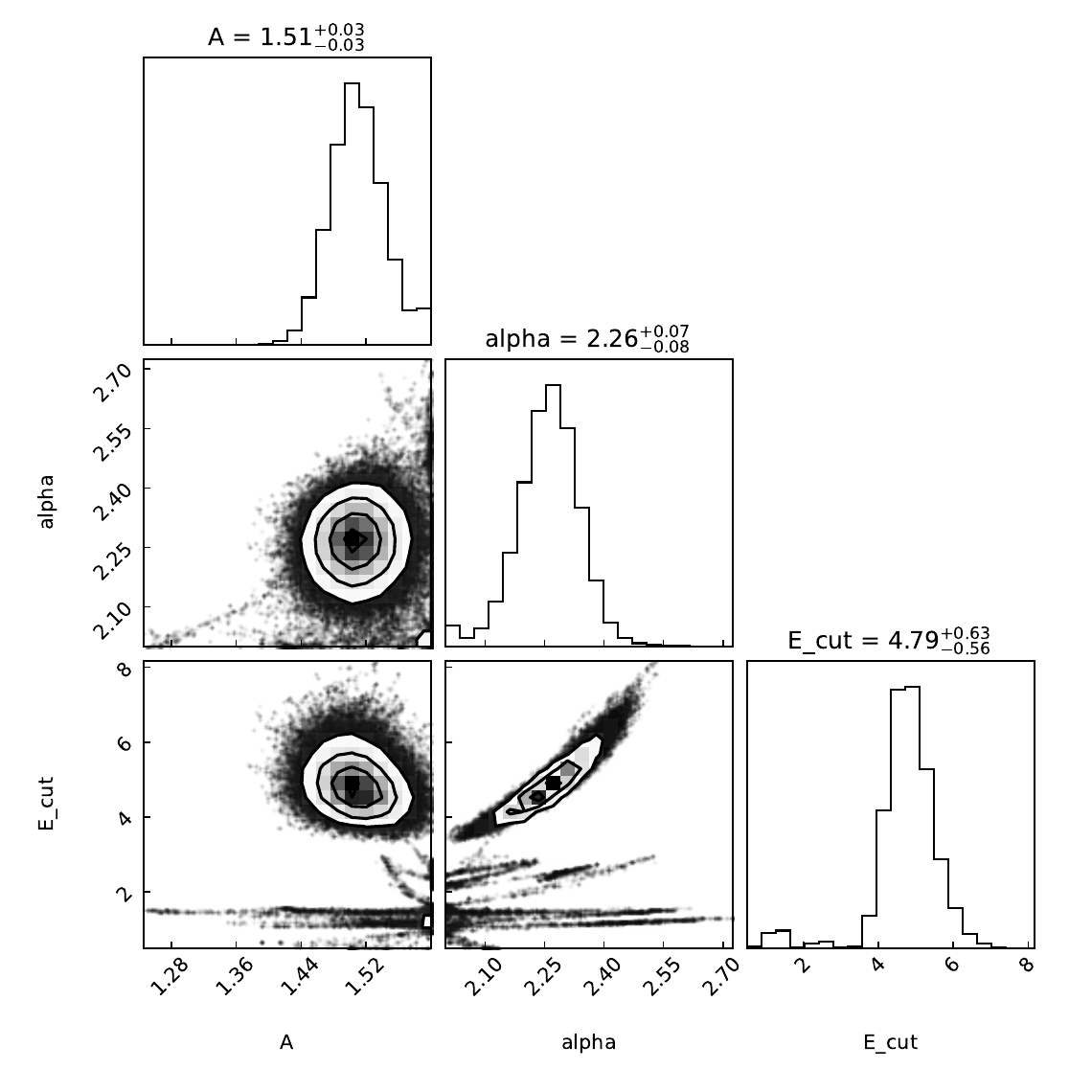}}
\subfigure[Mrk 501]{
\includegraphics[scale=0.45]{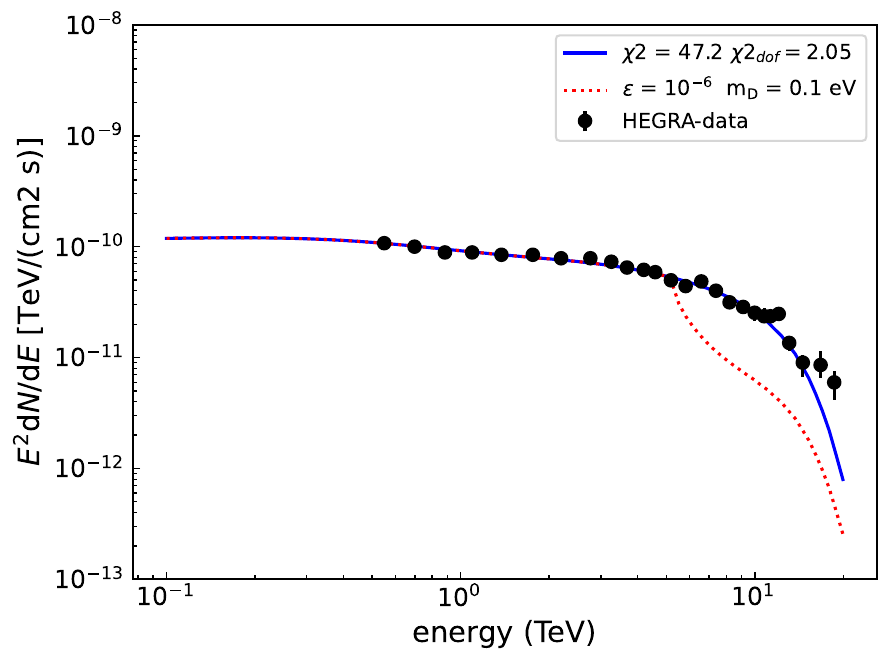}}
\subfigure[The posterior distribution of ECPL parameters for Mrk 501]{
\includegraphics[scale=0.27]{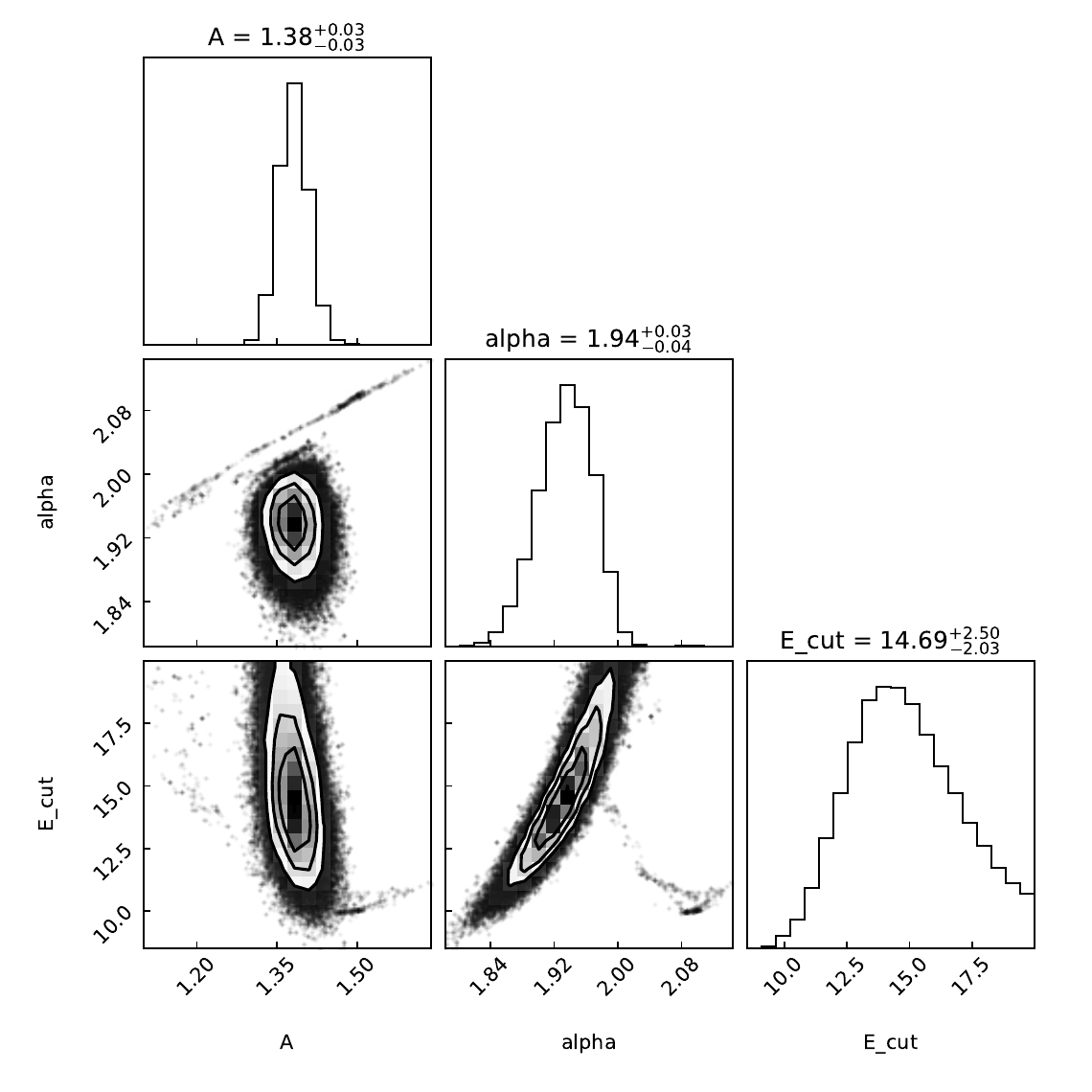}}
\caption{The HEGRA data are presented by the black points. 
The blue lines illustrate the best-fit energy spectra of (a) Mrk 421 and (c) Mrk 501 with the null hypothesis.
The red dashed lines present the effect from dark photon with $\epsilon=10^{-6}$ and $m_{\rm D}=0.1$ eV on the spectra of (a) Mrk 421 and (c) Mrk 501.
The right panels (b) and (d) present posterior distribution of ECPL parameters for Mrk 421 and Mrk 501 in MCMC runs respectively}.

\label{sed_fit}
\end{figure*}

\section{Method}
\label{sect:method}

The photon-dark photon scattering would produce an extra absorption on the spectra and suppress the flux with the photon energy higher than the threshold. Therefore we could distinguish the effect from DPDM and give a constraint spectrally.

Mrk 421 and Mrk 501 were the first two extragalactic objects identified as VHE gamma-ray emitters \cite{1996ApJ...456L..83Q,1997A&A...320L...5B}.
The energy spectra of these sources are taken from HEGRA \cite{2001A&A...366...62A,2002A&A...393...89A}, and one of which has been chosen for constraining the axionlike particles (ALPs) \cite{2021PhRvD.104h3014L}. 

An exponent cutoff power-law (ECPL) model \cite{2001A&A...366...62A,2002A&A...393...89A,2021PhRvD.104h3014L} is adopted to fit the observation of Mrk 421 and Mrk 501 as the form of the intrinsic energy spectra,
\begin{equation}
    dN/dE = A(E/{\rm{TeV}})^{-\alpha}e^{-E/E_{\rm {cut}}}.
\end{equation}
The VHE energy spectra could also be described as other spectral models like LogParabola \cite{2017A&A...600A..89H,2021PhLB..82136611C} and superexponential cutoff power law (SECPL) \cite{2021PhRvD.103h3003L}.  Since the fitting results and the final constraints on $\epsilon$ based on these models are very similar (see \ref{sec:details for models} for details), our benchmark results in the main text use the ECPL model.

The best-fit parameters could be determined by the minimized $\chi^2$ value, and the generalized $\chi^2$ could be expressed as
\begin{equation}
   \chi^2 = \sum_{i=0}^n \frac{(F_i(\theta) - F_{i,\rm{obs}})}{\sigma_i^2}
\end{equation}
where $i$ is the $i$th data, $\theta$ and $F_i(\theta)$ are the model parameters and model expected value, $F_{i,\rm{obs}}$ is the observed data and ${\sigma_i}$ is the observed data error.
We use the Markov Chain Monte Carlo (MCMC) method as a tool to find out the best-fit parameters and employ the $emcee$\footnote{\url{https://emcee.readthedocs.io/en/stable/index.html}} python package to realize MCMC.

The $\chi^2_{\rm {w/o}}$ under the null hypothesis could be defined as $\chi^2_{\rm {w/o}} = \sum_{i=0}^n \frac{(F_i(A,\alpha, E_{\rm {cut}}) - F_{i,\rm{obs}})}{\sigma_i^2}$, and the minimized $\chi^2_{\rm {w/o}}$ and corresponding fitting parameters are listed in Table \ref{without DP result}.
The HEGRA data and spectra with best-fit parameters are shown in Fig. \ref{sed_fit}, where the corner maps in the right panels are the posterior distributions of the null hypothesis in MCMC runs.
The red dashed lines represent the effect from DPDM on spectra for an example, where the photon - dark photon scattering would open up at the energy threshold $\sim 5$ TeV for 0.1 eV dark photon.

\begin{figure}[htbp]
\centering
\includegraphics[width=0.45\textwidth]{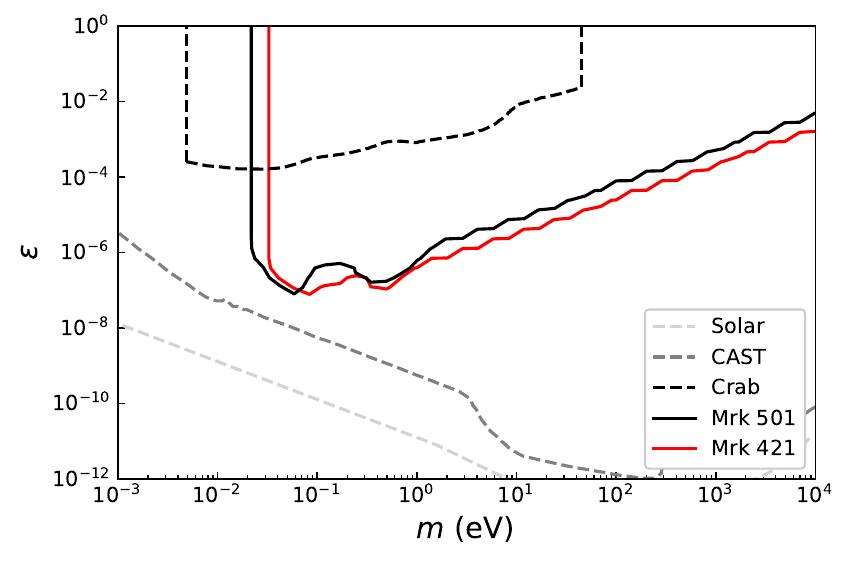}
\caption{Constraints on kinetic mixing from Mrk 501 (black) and Mrk 421 (red), which are represented as solid lines. While the comparing results are represented as dashed lines: the result from the CAST experiment by only considering the contribution from the transverse mode (grey) \cite{2010PhLB..689..149E}, the most stringent constraint by calculating the cooling rates for the Sun (light grey) \cite{2013PhLB..725..190A}, and a similar constraining approach with Crab Nebula (black) \cite{2021PhRvD.103d3018B}. }
\label{4Rs}
\end{figure}

\begin{figure*}[htbp]
\centering
\subfigure[]{
\includegraphics[width=0.45\textwidth]{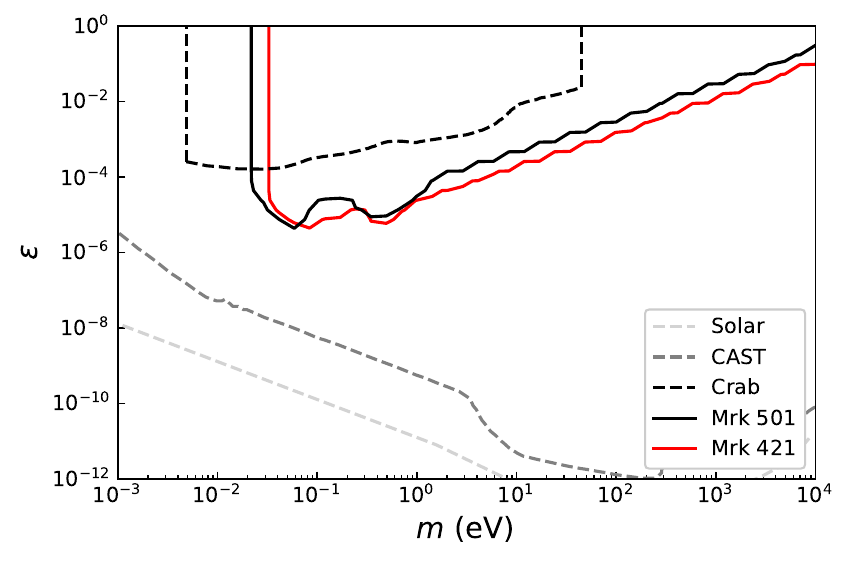}
}
\subfigure[]{
\includegraphics[width=0.45\textwidth]{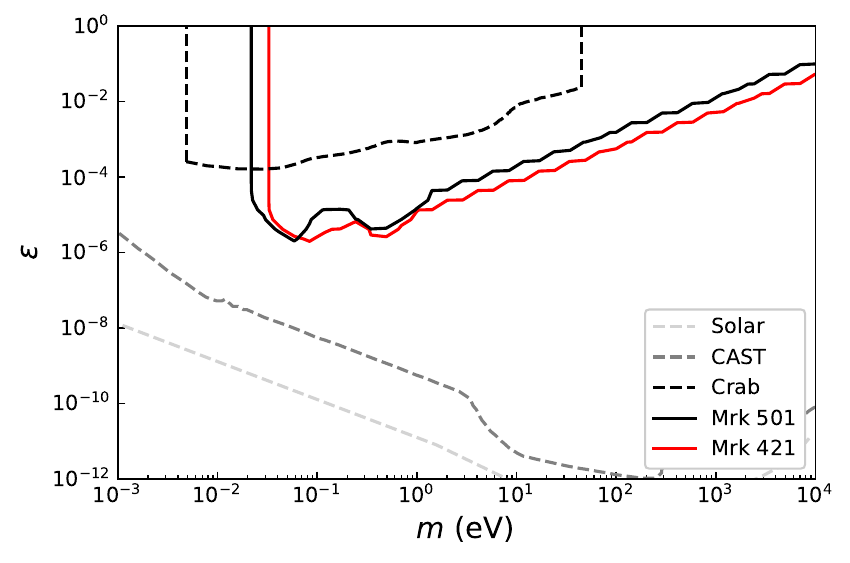}
}
\caption{Same as Fig. \ref{4Rs} but with $R_{\rm em}$=$10^4 R_s$ (left panel) and $\gamma_{\rm sp}$=1.5 (right panel).}
\label{conservative}
\end{figure*}

The spectral model with the DPDM considered could be expressed as
\begin{equation}
\label{DP}
    dN/dE = A(E/{\rm{TeV}})^{-\alpha}e^{-E/E\rm{_{cut}}}P_{\rm DP}.\end{equation}
The photon survival probability from dark photon $P_{\rm DP}$ includes the DPDM in IGM space, AGN host galaxy and the Milky Way, where the spike in the center of host galaxy dominates the $P_{\rm DP}$.

The parameter space $m_{\rm D}$ in (10$^{-2} - 10^4$) eV and $\ep$ in ($10^{-10} - 1$) are divided into a (40 $\times$ 40) gird with logarithmic steps. Each $P_{\rm DP}$ is calculated for a given set of DPDM parameters ($m_{\rm D}$, $\ep$).
The MCMC method is also employed to find out the minimized $\chi^2_{\rm {DP}}$ for each energy spectrum in each parameter grid, where $\chi^2_{\rm {DP}} = \sum_{i=0}^n \frac{(F_i(A,\alpha, E_{\rm {cut}},\ep,m_{\rm D}) - F_{i,\rm{obs}})}{\sigma_i^2}$.

Our strategy for constraining dark photon parameters is as follows:
The null hypothesis is the best-fitting result of energy spectra without DPDM, which corresponds to the minimized $\chi^2_{\rm w/o}$. 
In each DPDM parameter grid, we obtain the minimized $\chi^2_{\rm {DP}}$ of energy spectra by freeing the null hypothesis parameters, and calculate the $\Delta \chi^2$ = $\chi^2_{\rm DP}$ -  $\chi^2_{\rm w/o}$ to exclude that parameter region. 
This analysis method has been used for lots of works to constrain DM parameters from the effect on ordinary photon spectra \cite{2021PhRvD.103d3018B,2021PhLB..82136611C}.

The VHE photons from extragalactic sources on their way to the Earth would be scattered by extragalactic background light (EBL) \cite{2013APh....43..112D}. Therefore, we should take into account the absorption from the EBL effect when VHE photons propagate in the intergalactic medium (IGM). 
Several EBL models have been reported in the literature \cite{2011MNRAS.410.2556D, 2012MNRAS.422.3189G,2022ApJ...941...33F,2017A&A...603A..34F,2021MNRAS.507.5144S} and we employ the model in \cite{2011MNRAS.410.2556D} as the main result in this work.
Since the EBL effect is also an absorption, we test various EBL models to analyze the degeneracy with the absorption from DPDM in \ref{sec:details for EBL}. Different EBL models take little effect on constraining the DPDM parameter space.

\section{Results}
\label{sect:results}

A 95$\%$ confidence level (C.L.) exclusion region on kinetic mixing at a given dark photon mass $m_{\rm D}$ is derived in Fig. \ref{4Rs} corresponding to $\Delta \chi^2 \geq$ 2.71.  
The kinetic mixing is constrained to be $\ep \sim 10^{-7}$ at 95$\%$ C.L. with $m_{\rm D}\sim 0.03 - 1$ eV mass range by Mrk 421 and Mrk 501 based on the ECPL spectral model and the EBL model in \cite{2011MNRAS.410.2556D}, where $R_{\rm em}$ is adopted as the minimum spike region $4R_s$ \cite{2022PhRvL.128v1104W}.

The result of Mrk 501 could be close to that from CAST \cite{2010PhLB..689..149E} in a narrow mass range but still much weaker than the most stringent constraint in \cite{2013PhLB..725..190A}. A previous result of the Crab Nebula with a similar approach is also plotted to be contrasted \cite{2021PhRvD.103d3018B}. The Crab Nebula is a well-known VHE source with a sub-PeV gamma-ray spectrum observed by the Tibet AS$\gamma$ experiment \cite{2019PhRvL.123e1101A}. That much higher photon energy could give a large exclusion region with lighter dark photons. 
However, the absorption from cosmic microwave background (CMB) and EBL are enormous so that a blazar with above 100 TeV photons is hardly detected.

We also plot the results that $R_{\rm em}$ = $10^4 R_s$ in Fig. \ref{conservative} left panel as a conservative situation. In that extreme case, the VHE photons from the blazar jet are assumed to be produced at the edge of the spike and could hardly pass through the most dense spike region. Therefore, the absorption from the spike is deeply suppressed and is comparable with that from a canonical NFW profile.

The results of the Gnedin and Primack spike are also shown in Fig. \ref{conservative} right panel as another conservative situation. $R_{\rm em}$ is also adopted to be $10^4 R_s$. Note that $R_{\rm em}$ has little effect on the constraining results since the cusp index of the spike is too low to dominate the DM integral. In that case, a much larger propagating distance in the NFW profile would counteract the influence of dense DM density of the spike.

\begin{figure}
\centering
\includegraphics[scale=0.55]{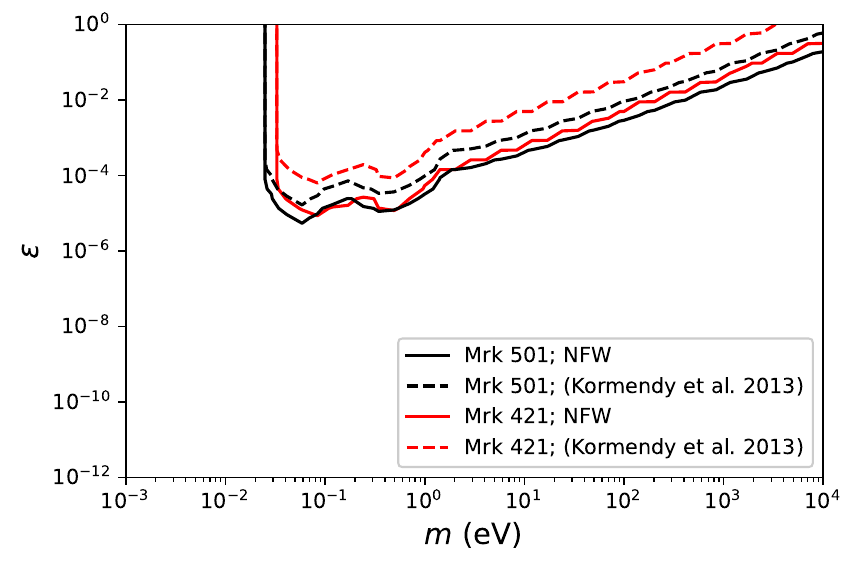}
\caption{The constraints on DPDM from Mrk 421 (red line) and Mrk 501 (black line) are obtained by adopting the host galaxies as NFW profile. The results with another estimation method is represented as the dashed line.}
\label{NFW results}
\end{figure}

The existence of DM spikes is still unconfirmed since there is no evidence from observations or N-body simulations. 
It is necessary for a seed black hole of SMBH to grow on a timescale long compared with $10^7$ years, and this process should occur within $50$ pc of the center of DM halo composed of very cold DM particles. 
The time needed for the adiabatic growth is even shorter for high-redshift AGNs that we considered.  
Although a black hole spiraled into the potential well of DM halo from outside could also form a DM spike, the spiral-in timescale would be roughly comparable to the Hubble time depending on the black hole mass and distance \cite{2001PhRvD..64d3504U}. 
These strict forming conditions would also lead to a significant uncertainty on the spike formation.
Therefore, we give the constraining results in Fig. \ref{NFW results} with the spike removed.

Since the normalization of NFW profile at blazar host galaxies is calculated from \ref{normalization} where the existence of DM spike is assumed, we also try to use another method to estimate that normalization.
The correlation between the SMBH and total DM halo masses of AGNs is shown in \cite{2013ARA&A..51..511K} by assuming the bulge and stellar masses are almost the same. 
We use a middle value of the correlation to estimate the total DM halo masses and the normalization of NFW profile of Mrk 421 and Mrk 501 host galaxies. The constraining results are also shown in Fig. \ref{NFW results} to be contrasted.

\section{Conclusion}
\label{sect:conclusion}

In this work, we constrain the kinetic mixing over photon-dark photon scattering. The VHE photons from the relativistic jet of blazar would pass through the DM spike and be strongly influenced by the dense DPDM. 
Besides, VHE photons could give the constraining region on low DPDM mass because of the scattering energy threshold. 
Therefore, blazars Mrk 421 and Mrk 501 are suitable sources to constrain the kinetic mixing.

DPDM in IGM space and the Milky Way halo are calculated though, the part of DM spike surrounding SMBH in the center of the blazar host galaxy dominates the absorption of VHE photons. An extreme situation that photons induced at the edge of the spike could hardly pass through. In that case, the absorption would be deeply suppressed. Another conservative situation that the Gnedin and Primack spike where $\gamma_{sp}$ = 1.5 is also discussed.
The kinetic mixing of DPDM is constrained to be $\ep \sim 10^{-7}$ with $m_{\rm D}\sim 0.03 - 1$ eV mass range in our best result and provide a validation to other constraints.

Mrk 501 gives a better constraint than Mrk 421 in a narrow mass range of $\sim$ 0.02-0.03 eV since its higher photon energy. Gamma-Ray Burst (GRB) 221009A (known as the most energetic GRB ever detected) might also provide a chance to constrain the DPDM parameters \cite{2023arXiv230606372L,2023arXiv231008845C}.
However, GRB 221009A is close to the nucleus region of the host galaxy with a distance of 0.65 kpc though \cite{2023ApJ...946L..28L}, the photons still cannot be affected by the DM spike.

\section*{Acknowledgements}

We gratefully thank Yun-Feng Liang, Ting-Ting Ge, and Jin-Wei Wang for their useful discussions. This work is supported by the National Key Research and Development Program of China (Grant No. 2022YFF0503304), the Programme of Bagui Scholars Programme (WXG), the Guangxi Talent Program (“Highland of Innovation Talents”), and Innovation Project of Guangxi Graduate Education (YCBZ2024060).
\label{lastpage}

\appendix

\section{Different models of energy spectra}
\label{sec:details for models}

In order to test the reliability of the ECPL model, we adopt some other widely-used models to describe the energy spectra of Mrk 421 and Mrk 501, such as SECPL,
\begin{equation}
\label{SECPL}
    dN/dE = A(E/{\rm{TeV}})^{-\alpha}e^{-(E/E_{\rm {cut}})^d},
\end{equation}
and LogParabola,
\begin{equation}
\label{LogParabola}
    dN/dE = A(E/{\rm{TeV}})^{-(\alpha +\beta \log(E) )}.
\end{equation}

\begin{figure*}[htbp]
\centering
\subfigure[Mrk 421]{
\includegraphics[width=0.45\textwidth]{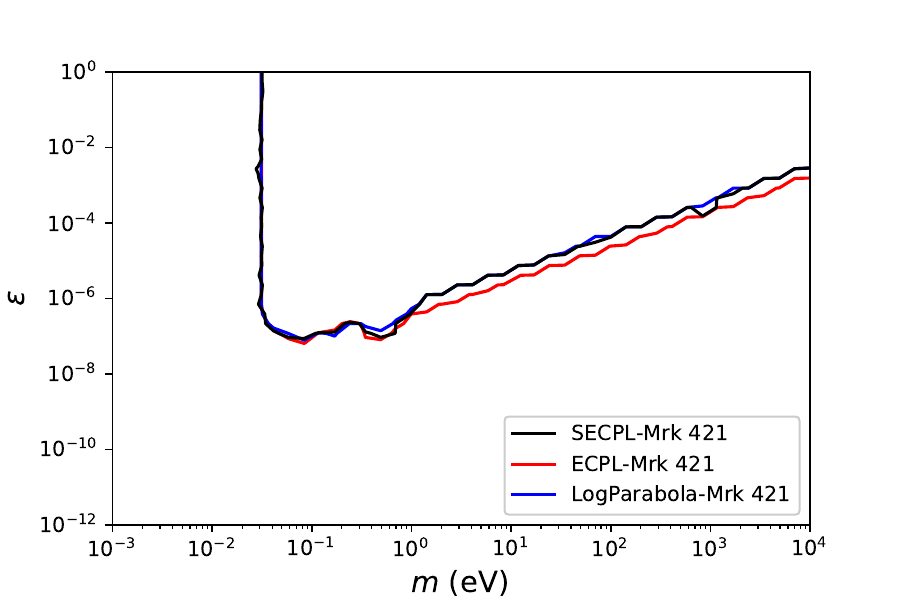}
}
\subfigure[Mrk 501]{
\includegraphics[width=0.45\textwidth]{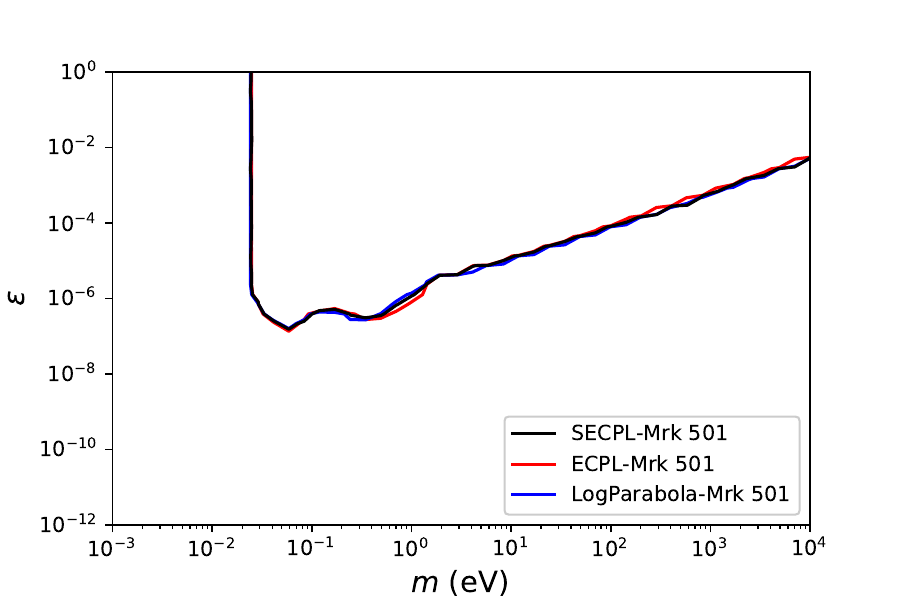}
}
\caption{The DPDM constraining results with ECPL (red), SECPL (black) and LogParabola (blue) models. The solid and dashed line represent the sources Mrk 421 and Mrk 501 respectively.}
\label{degeneracy test for null hypotheses}
\end{figure*}

\begin{figure*}[htbp]
\centering
\subfigure[]{
\includegraphics[scale=0.45]{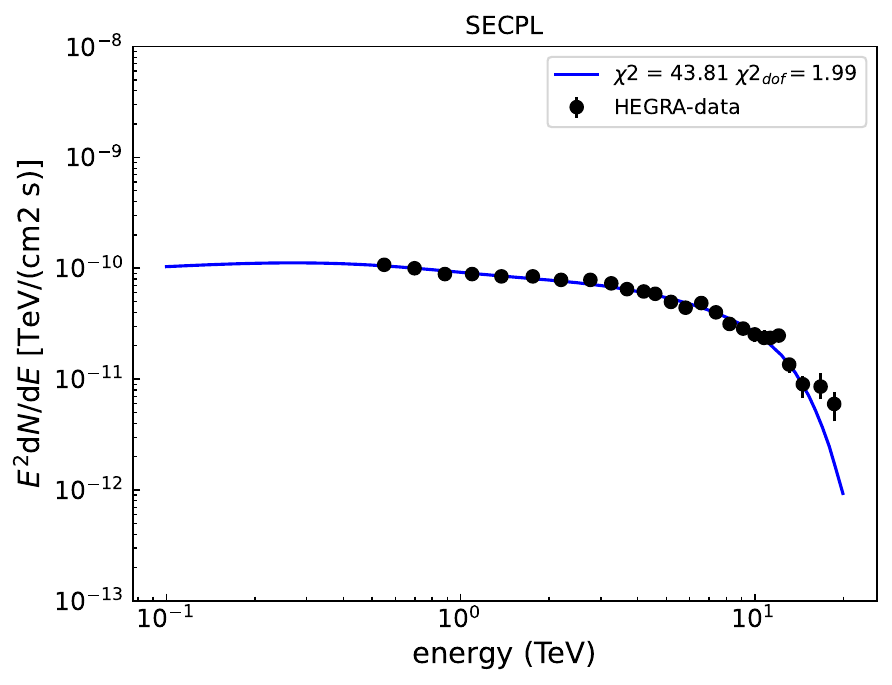}
}
\subfigure[]{
\includegraphics[scale=0.21]{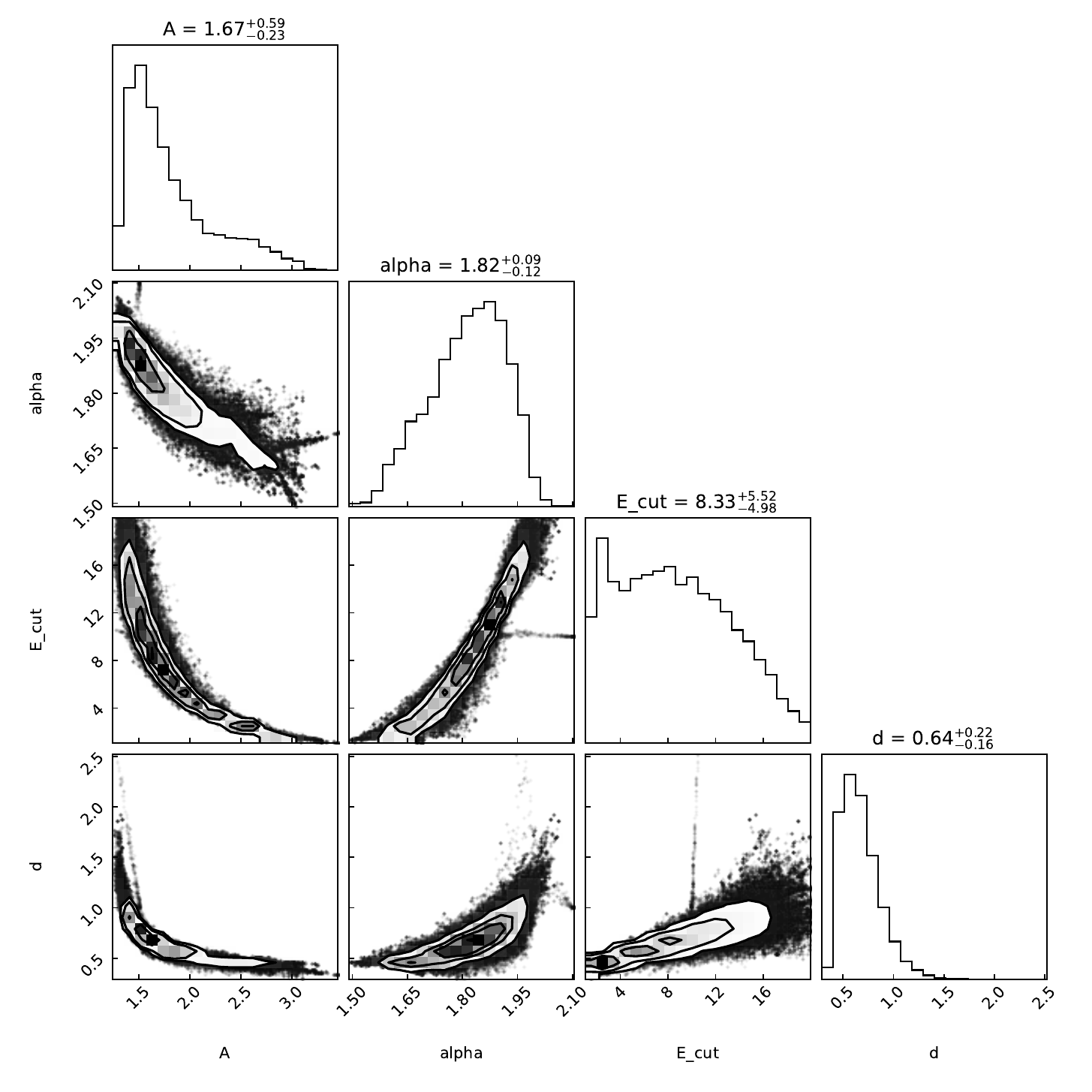}
}
\subfigure[]{
\includegraphics[scale=0.45]{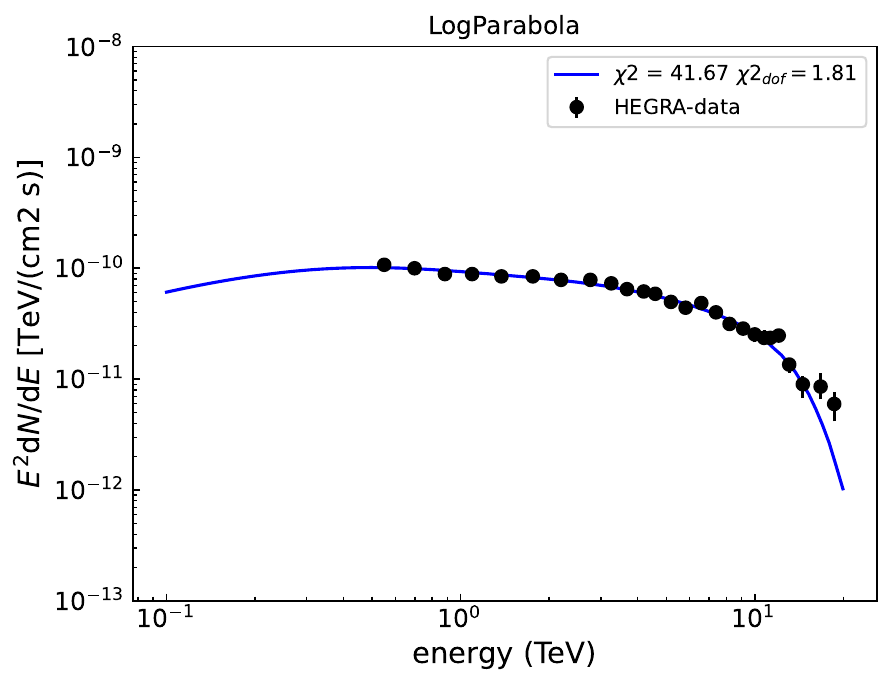}
}
\subfigure[]{
\includegraphics[scale=0.27]{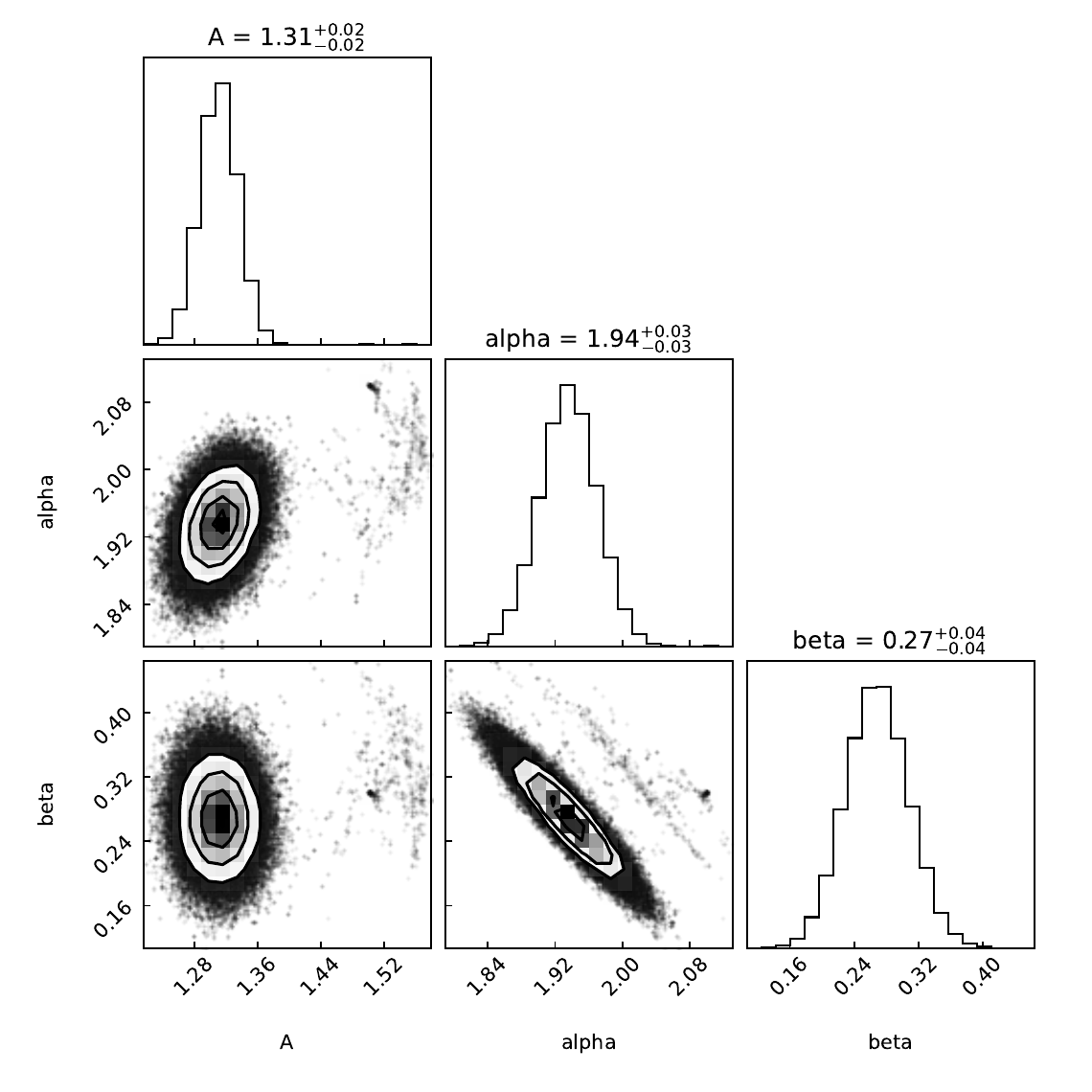}
}
\caption{The best fitting results (left panels) and posterior distribution in MCMC runs (right panels) for Mrk 501 with the SECPL and LogParabola model.}
\label{Mrk 501 different null hypothesis}
\end{figure*}

\begin{figure*}[htbp]
\centering
\subfigure[]{
\includegraphics[width=0.45\textwidth]{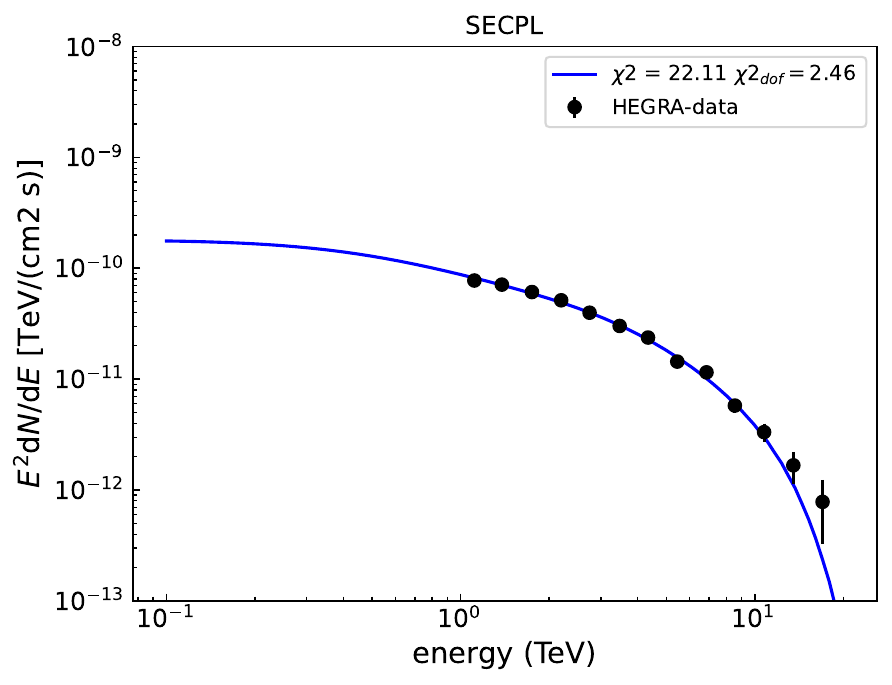}
}
\subfigure[]{
\includegraphics[width=0.34\textwidth]{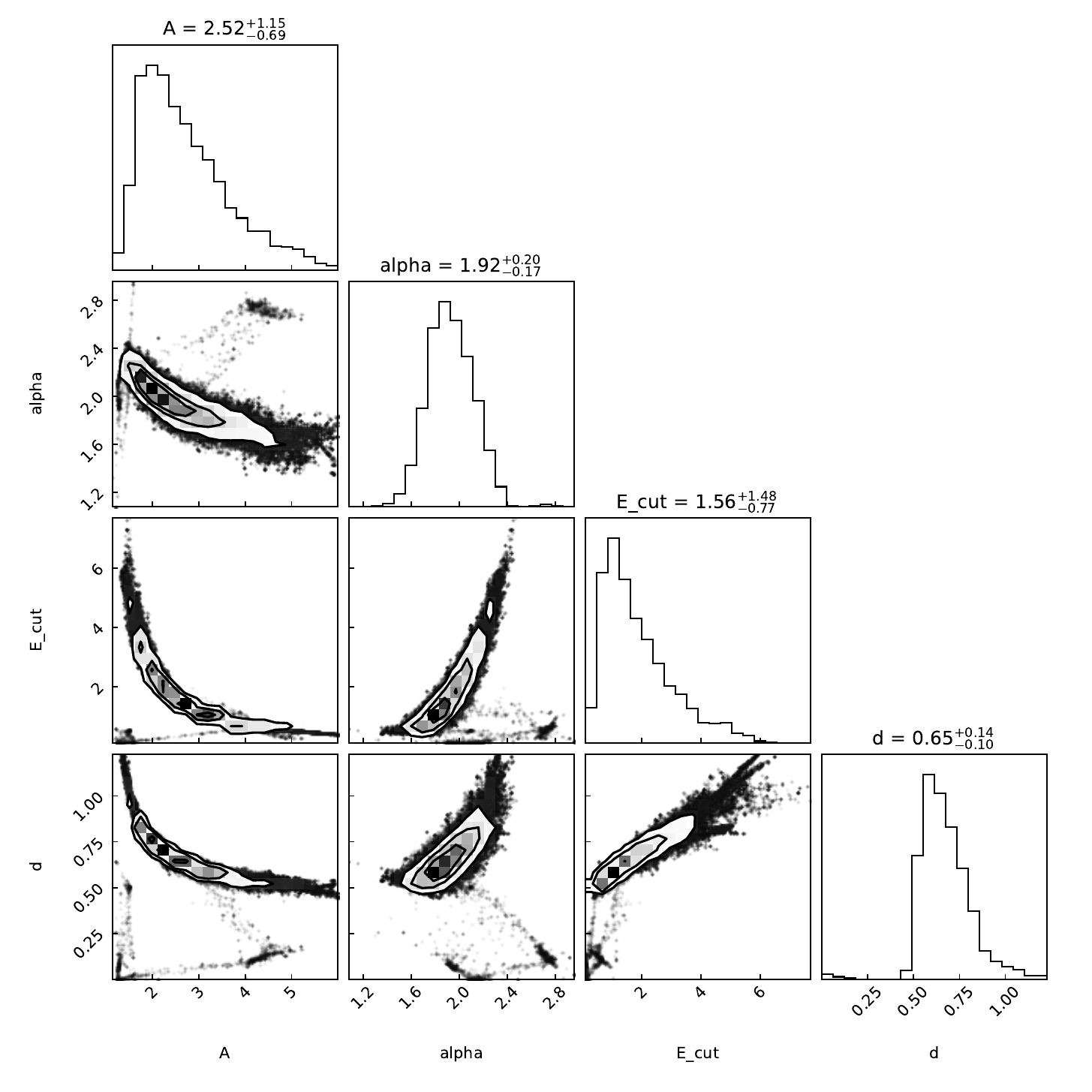}
}
\subfigure[]{
\includegraphics[width=0.45\textwidth]{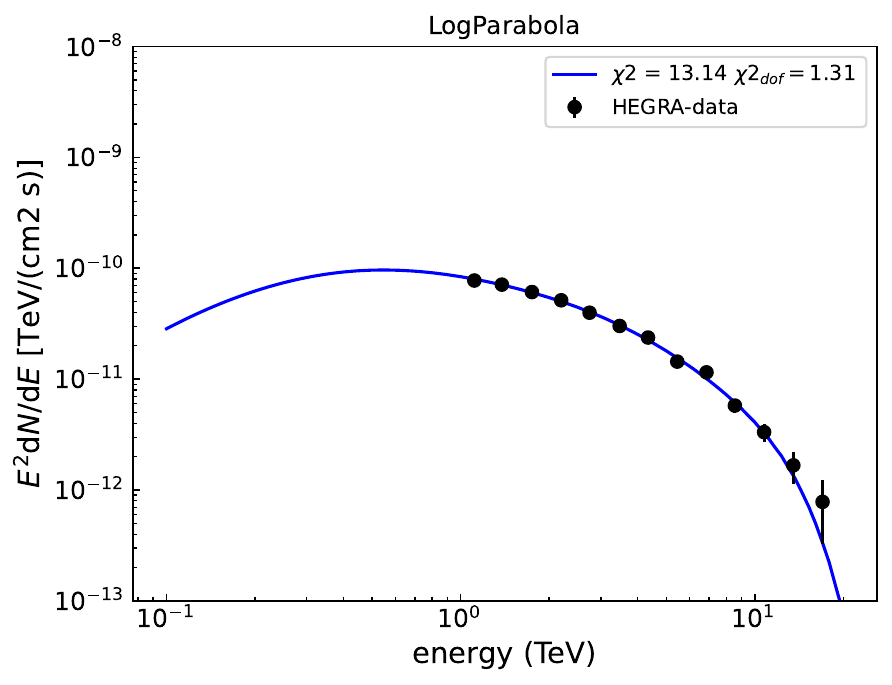}
}
\subfigure[]{
\includegraphics[width=0.34\textwidth]{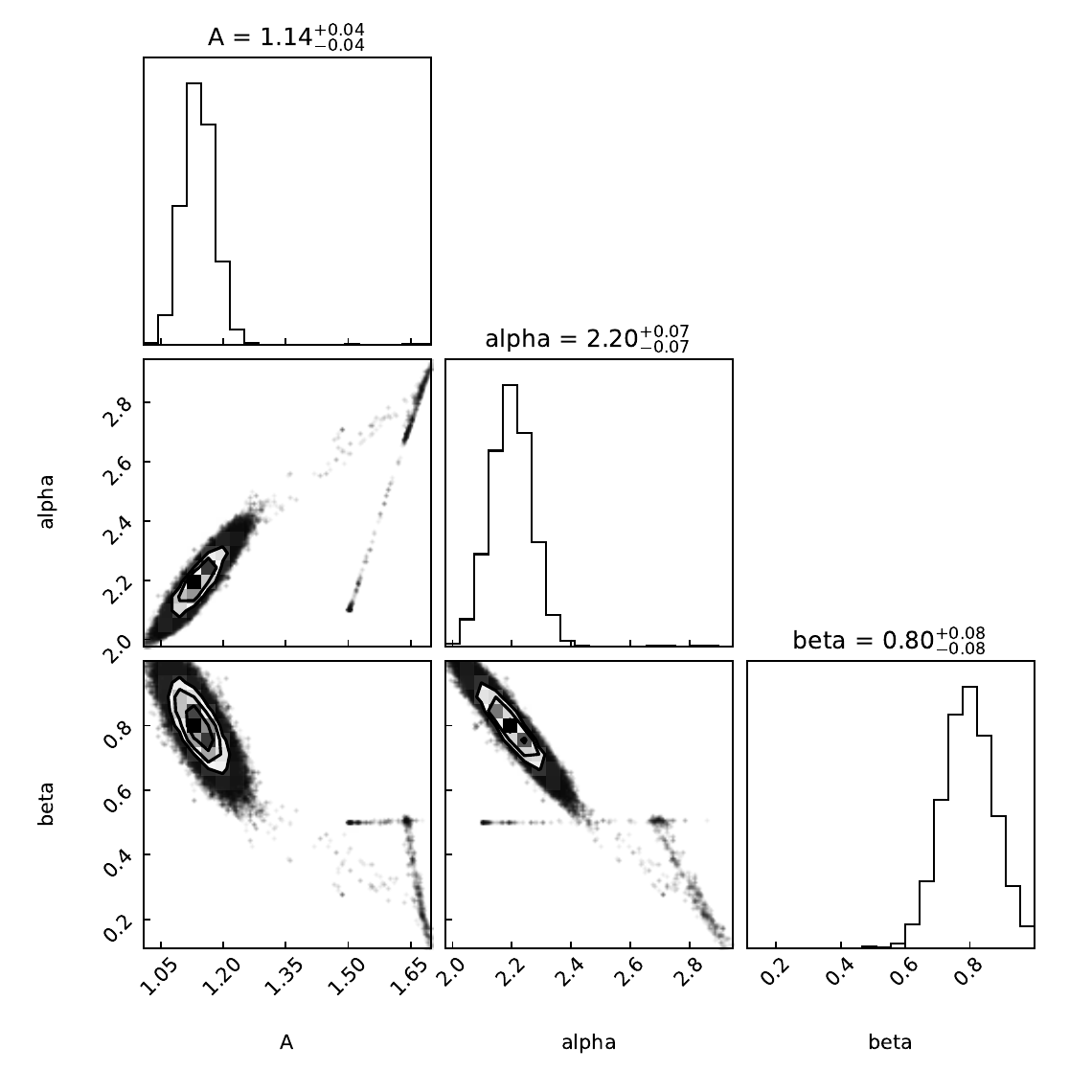}
}
\caption{The best fitting results (left panels) and posterior distribution in MCMC runs (right panels) for Mrk 421 with the SECPL and LogParabola model.}
\label{Mrk 421 different null hypothesis}
\end{figure*}

The prior ranges of the paramaters and the minimized $\chi^2$ of different spectral models are listed in Table \ref{different model}.
The best fitting results and posterior distribution in MCMC runs are plotted for Mrk 501 (Fig. \ref{Mrk 501 different null hypothesis}) and Mrk 421 (Fig .\ref{Mrk 421 different null hypothesis}).
The final constraining results with different models are also shown in Fig. \ref{degeneracy test for null hypotheses}.

Overall, the energy spectra of Mrk 501 and Mrk 421 could be well described by all of the spectral models we considered. The constraining results from the models are very similar. Therefore we choose to present our main results with ECPL model.

\begin{table*}
	\begin{center}
	\caption{The prior ranges of parameters and the minimized $\chi^2$ with different null-hypothesis model.}
	\label{different model}
	\renewcommand\arraystretch{1.5}
	\begin{tabular}{l|ccccccc}
	\hline
	Source &   model & prior range &&&&& $\chi^{2}/dof$\\
        & &A ($10^{-10}$TeV$^{-1}$cm$^{-2}$s$^{-1}$) & $\alpha$ & $E_{\rm {cut}}$&$\beta$&d&\\
	\midrule
    Mrk 501 & ECPL & [0.5, 6] & [1, 5] & [1, 20]&-& - & 47.23/23\\
    Mrk 501 & SECPL & [0.5, 6] & [1, 5] & [1, 20]& -& [0, 3]&43.81/22\\
    Mrk 501 & LogParabola & [0.5, 6] & [1, 5] & -& [0.1, 0.5]& -& 41.67/23\\
    
    Mrk 421 & ECPL & [0.5, 6] & [1, 5] & [1, 20]& - & - & 24.39/10\\
    
    Mrk 421 & SECPL & [0.5, 6] & [1, 5] & [0.1, 20]& - & [0, 3]&22.11/9\\
    
    
    Mrk 421 & LogParabola & [0.5, 6] & [1, 5] & -& [0.1, 1]& -& 13.14/10\\
	
	\hline
	\end{tabular}
	\end{center}
\end{table*}

\section{Different EBL models}
\label{sec:details for EBL}

\begin{figure*}[htbp]
\centering
\subfigure[Mrk 421]{
\includegraphics[width=0.45\textwidth]{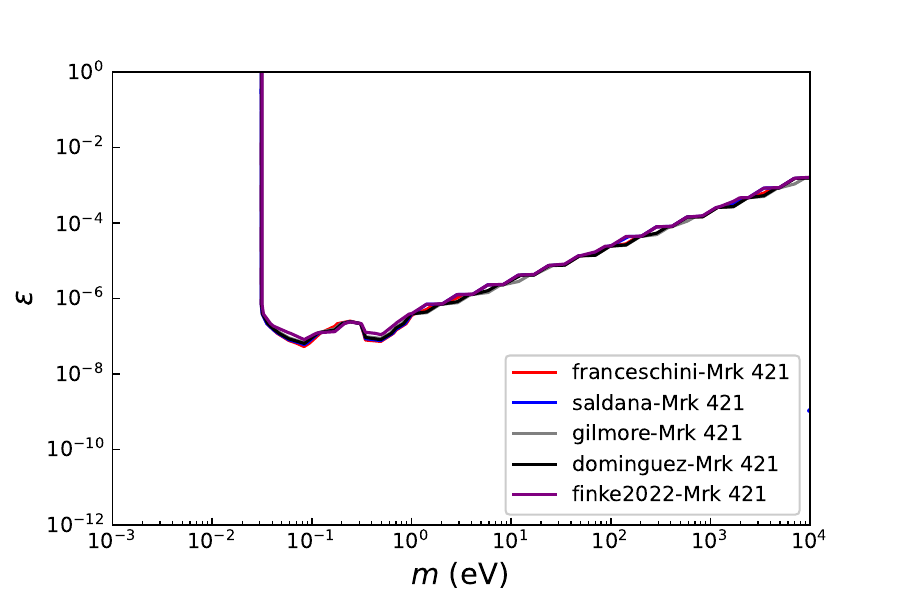}
}
\subfigure[Mrk 501]{
\includegraphics[width=0.45\textwidth]{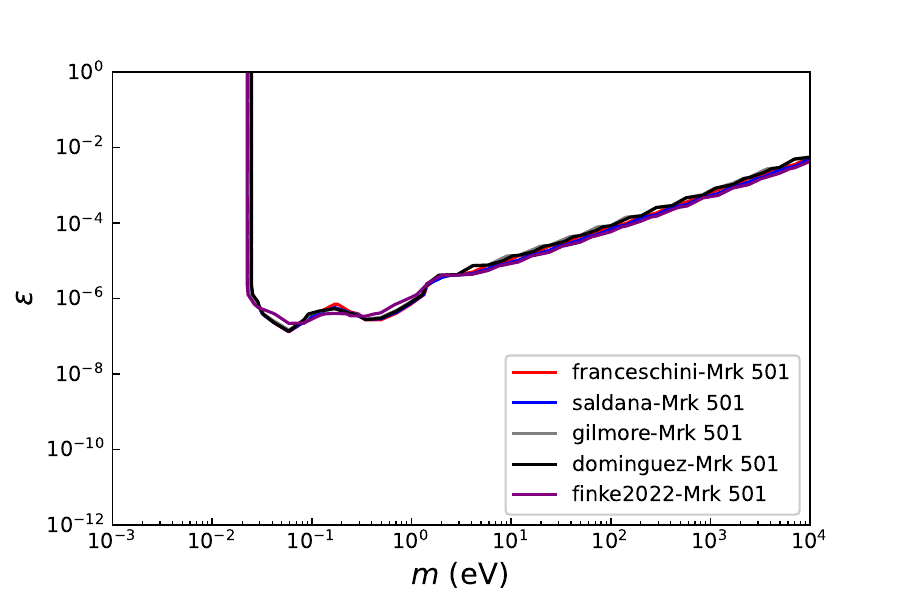}
}
\caption{The DPDM constraining results with different EBL models for (a) Mrk 421 and (b) Mrk 501.}
\label{degeneracy test for EBL}
\end{figure*}

Since the VHE photon from blazars is able to be absorbed by both EBL photon and dark photon, there could be some degeneracy exists to have an impact on the constraining results.
The DPDM constraints with various EBL models are illustrated in Fig. \ref{degeneracy test for EBL}. We could see that the final results are very similar, and thus we adopt the EBL model in \cite{2011MNRAS.410.2556D} as the main result.

Besides, other effects that would be degenerate with the absorption from DPDM are also considered. However, the energy of CMB photon is too low to absorb the VHE photon, and the Galactic background filed deeply rely on the position information of the source \cite{2015JCAP...10..014E}, which means it can hardly have an impact on Mrk 421 and Mrk 501 for their high galactic latitude.

%
\clearpage
\bibliographystyle{spphys.bst}
\bibliography{spphys.bib}


\end{document}